\documentclass[prd,twocolumn,showpacs,preprintnumbers,amsmath,nofootinbib,amssymb,floatfix]{revtex4-1}
\usepackage{graphicx}
\usepackage[sort&compress]{natbib}
\usepackage{subfigure}
\usepackage{amsmath}
\usepackage{amsfonts}
\usepackage{cancel}
\usepackage{lmodern,dsfont}
\usepackage{amssymb}
\begin{document}
\arraycolsep1.5pt
\newcommand{\Ima}{\textrm{Im}}
\newcommand{\Rea}{\textrm{Re}}
\newcommand{\mev}{\textrm{ MeV}}
\newcommand{\be}{\begin{equation}}
\newcommand{\ee}{\end{equation}}
\newcommand{\ba}{\begin{eqnarray}}
\newcommand{\ea}{\end{eqnarray}}
\newcommand{\gev}{\textrm{ GeV}}
\newcommand{\nn}{{\nonumber}}
\newcommand{\dtres}{d^{\hspace{0.1mm} 3}\hspace{-0.5mm}}
\newcommand{\rts}{ \sqrt s}
\newcommand{\non}{\nonumber \\[2mm]}

\title{Strategy to find the two $\Lambda(1405)$ states from lattice QCD simulations}

\author{A. Mart\'inez Torres$^1$, M. Bayar$^{2,3}$, D. Jido$^{1,4}$ and E. Oset$^2$}
\affiliation{
$^1$Yukawa Institute for Theoretical Physics, Kyoto University, Kyoto 606-8502, Japan.\\
$^2$Departamento de F\'{\i}sica Te\'orica and IFIC, Centro Mixto Universidad de Valencia-CSIC,
Institutos de Investigaci\'on de Paterna, Aptdo. 22085, 46071 Valencia, Spain.\\
$^3$ Department of Physics, Kocaeli University, 41380 Izmit, Turkey.\\
$^4$ J-PARC Branch, KEK Theory Center,
Institute of Particle and Nuclear Studies, 
High Energy Accelerator Research Organization (KEK),
203-1, Shirakata, Tokai, Ibaraki, 319-1106, Japan.
}

\date{\today}

\begin{abstract}

  Theoretical studies within the chiral unitary approach, and recent experiments, have provided evidence of the existence of two isoscalar states in the region of the $\Lambda(1405)$. In this paper we use the same chiral approach to generate energy levels in a finite box. In a second step, assuming that these energies correspond to lattice QCD results, we devise the best strategy of analysis to obtain the two states in the infinite volume case, with sufficient precision to distinguish them. We find out that by using energy levels obtained with asymmetric boxes and/or with a moving frame, with  reasonable errors in the energies, one has a successful scheme to get the two $\Lambda(1405)$ poles.

\end{abstract}
\preprint{YITP-12-6}
\preprint{J-PARC-TH 007}
\maketitle

\section{Introduction}
\label{Intro}

   The history of the $\Lambda(1405)$ as a composite state of meson baryon, dynamically generated from the meson baryon interaction, is rather long, starting from the works of  Refs.~\cite{Dalitz:1960du,Dalitz:1967fp}. Early works using the cloudy bag model also reached similar conclusions \cite{Veit:1984an}. The advent of chiral unitary theory, combining chiral dynamics and unitarity in coupled channels, brought new light onto this issue and the $\Lambda(1405)$ was one of the cleanest examples of states dynamically generated within this approach 
\cite{Kaiser:1995eg,Kaiser:1996js,angelskaon}. Hints that there could be two states rather than one had also been reported using the cloudy bag model   \cite{Fink:1989uk} and the chiral unitary approach \cite{ollerulf,Jido:2002yz,GarciaRecio:2002td}. A qualitative step forward was done in Ref.~\cite{cola}, where two different versions of the approach were used, the two poles remained, and their origin was investigated. It was found that in an SU(3) symmetric theory there were two degenerate octets and a singlet of dynamically generated resonances, but with the breaking of SU(3) the degeneracy was removed, one octet with isospin $I=0$ moved to become the $\Lambda(1670)$ and the other one moved close to the singlet, producing two poles close by in the region of the $\Lambda(1405)$. One of the poles appears at energies around 1420 MeV, couples mostly to $\bar K N$ and has a small width of around 30 MeV. The other pole is around 1395 MeV, couples mostly to $\pi \Sigma$ and is much wider, around 120 or 250 MeV depending on the model. After the work of Ref.~\cite{cola}, all further works on the chiral unitary approach have corroborated the two poles, with remarkable agreement for the pole at higher energy and larger variations for the pole at lower energies  \cite{carmenjuan,hyodo,Borasoy:2005ie,Oller:2005ig,Oller:2006jw,Borasoy:2006sr,Hyodo:2008xr,Roca:2008kr}. 

   Suggestions of experiments to confirm this finding were made, and it was shown that one should not expect to see two peaks in the cross sections, but rather different shapes in different reactions. In this sense, a suggestion was made to look for the  $\Lambda(1405)$ peak in the $K^- p \to \gamma \pi \Sigma$ reaction \cite{Nacher:1999ni}, where the $\gamma$ would be radiated from the initial state, making the $K^- p$ system lose energy and go below threshold and then excite the high energy state of the $\Lambda(1405)$, to which it couples most strongly. This reaction was not made, although it is planned for JPARC \cite{hayano}, but a similar one, where the photon was substituted by a pion, was implemented in 
Ref.~\cite{Prakhov:2004an} studying the $K^- p \to \pi^0 \pi^0 \Sigma^0$ reaction at $p_K = 514$~MeV/c - $750$ ~MeV/c. A neat and narrow peak was seen at $\sqrt s= 1420$~MeV, which was analyzed in Ref.~\cite{magas} and interpreted in terms of the high energy pole of the $\Lambda(1405)$.  More recently it was noticed that old data on the $K^- d \to \pi \Sigma ~n$ reaction from  Ref.~\cite{Braun:1977wd} produced a peak in the $\pi \Sigma$ spectrum around $\sqrt s= 1420$~MeV, with also a small width. These data were well reproduced in Ref.~\cite{sekihara} within the chiral unitary approach and multiple scattering, and once again it was shown that it gave support to the existence of the second pole of the $\Lambda(1405)$. It was shown there that the reaction proceeded with kaons in flight but not for stopped kaons, because the background from single scattering was too large in this latter case, obscuring the signal of the resonance that stems from double scattering. Even then, it was shown in Ref.~\cite{dafne} that kaons from the DAFNE facility, coming from the decay of the $\phi$, would also be suited to search for this resonance if neutrons were measured in coincidence in order to reduce the background. Results on the helicity amplitudes of the $\Lambda(1405)$  are also consistent with the two pole scenario \cite{mishajido}. The search for reactions where the  $\Lambda(1405)$ is produced has continued, showing that, as predicted, different reactions have different shapes. In this sense there have been recent photoproduction experiments \cite{Niiyama:2008rt,Moriya:2009mx} and proton induced experiments \cite{Zychor:2007gf,fabbietti} where the shapes are indeed different and the peaks appear at lower energies, around 1405 MeV, as the nominal mass.  There are also theoretical studies for these reactions where the peaks appear around these energies, and the larger contribution of the lower energy state that couples mostly to $\pi \Sigma$ is mostly responsible for it \cite{Nacher:1998mi,Hyodo:2003jw,Hyodo:2004vt,Nam:2008jy,Geng:2007vm}.

In as much as chiral dynamics is a good representation of QCD at low energies, the predictions of the chiral unitary approach on the  $\Lambda(1405)$ stand on firm ground. Yet, it would also be very interesting to have these predictions confirmed with lattice QCD simulations. In this sense,  
 the determination of hadron spectra is one of the challenging tasks
of Lattice QCD and many efforts are being devoted to this problem
\cite{Nakahara:1999vy,Mathur:2006bs,Basak:2007kj,Bulava:2010yg,Morningstar:2010ae,Foley:2010te,Alford:2000mm,Kunihiro:2003yj,Suganuma:2005ds,Hart:2006ps,Wada:2007cp,Prelovsek:2010gm,Lin:2008pr,Gattringer:2008vj,Engel:2010my,Mahbub:2010me,Edwards:2011jj,Lang:2011mn,Prelovsek:2011im},  some of them in particular to the search of the $\Lambda(1405)$ \cite{Melnitchouk,Nemoto,fxlee,TBurch,ishii, TTTakahashi}. A review on the $\Lambda(1405)$ and attempts to see it from different points of view is given in Ref.~\cite{jidohyodorev}.  
In some works the  ``avoided level crossing'' is usually taken as a signal of a
resonance, but this criteria has been shown insufficient for resonances
with a large width
\cite{Bernard:2007cm,Bernard:2008ax,misha}. Sometimes, the lattice spectra at finite volumes is directly associated to the energies of the states in infinite volume invoking a weak volume dependence of the results, as done recently searching for the $\Lambda(1405)$ resonance \cite{mena}. A more
accurate method consists on the use of L\"uscher's approach, for resonances with one decay channel. The method allow us to reproduce the phase shifts
for the decay channel from the discrete energy levels in the box
\cite{luscher,Luscher:1990ux}. This method has been recently simplified and improved in
Ref.~\cite{misha} by keeping the full relativistic two body propagator
(L\"uscher's approach keeps the imaginary part of this propagator exactly but makes approximations for the real part) and extending the method to two or
more coupled channels. The method has also been applied in
Ref.~\cite{mishajuelich} to obtain finite volume results from the J\"ulich model for
meson baryon interaction, including spectra for the $\Lambda(1405)$ with finite volume, and in Ref.~\cite{alberto}, to study the interaction of
the $DK$ and $\eta D_s$ system where the $D_{s^*0}(2317)$ resonance is
dynamically generated from the interaction of these particles
\cite{Kolomeitsev:2003ac,Hofmann:2003je,Guo:2006fu,daniel}. The case of
the $\kappa$ resonance in the $K \pi$ channel is also addressed along
the lines of Ref.~\cite{misha} in Ref.~\cite{mishakappa}. A first attempt to get phase shifts and the position of the $\Lambda(1405)$ from pseudo lattice data is done in Ref.~\cite{ulfnew}, where a different method is suggested and a qualitative study is made on how it could work.

   In the work of Ref.~\cite{misha}, the inverse problem of getting phase shifts and resonances from lattice results using two channels was addressed, paying special attention to the evaluation of errors and the precision needed on the lattice results to obtain phase shifts and resonance properties with a desired accuracy. Further work along these lines is done in Ref.~\cite{mishakappa}. The main problem encountered is that the levels obtained from the box of a certain size range do not cover all the desired energy region that one would like to investigate. Several suggestions are given in order to produce extra levels, like using twisted boundary conditions or asymmetric boxes \cite{misha}. These are, however, not free of problems since it is unclear whether a full twisting can be done in actual QCD simulations including sea quarks, and the asymmetric boxes have the problem of the possible mixing of different partial waves. Another alternative is to evaluate levels for a system in a moving frame as done in 
Ref.~\cite{Lang:2011mn}, but this also poses problems of mixing in principle. The generalization of  L\"uscher's approach to the moving frame is done in 
Refs.~\cite{Rummukainen:1995vs,sachraj,arXiv:1108.5371,arXiv:1110.0319,mishamoving}, and it provides a convenient framework for lattice calculations since new levels can be obtained without enlarging the size of the box, with an economy in computational time. It is then quite convenient to carry out simulations using effective theories in a finite volume, preparing the grounds for future lattice calculations, trying to find an optimal strategy on which configurations to evaluate in order to obtain the desired observables in the infinite volume case. 
  
  The case of extracting the $\Lambda(1405)$ parameters is specially challenging, particularly because two resonance must be found which are not too far from each other, which means that extra precision will be demanded of the lattice results. Furthermore, the two poles are not to be seen in the $\pi \Sigma$ phase shifts, since, as mentioned before, different amplitudes give different weight to the two poles and the $\pi \Sigma$ phase shifts provide insufficient information. The other reason is that the chiral unitary approach tells us that the two states couple strongly to $\bar K N $ and $\pi \Sigma$, so the use of the two channels in the analysis is mandatory and the use of one channel as in the L\"uscher approach is bound to produce incorrect results. 
In view of this we face the problem using the two channels explicitly in the analysis and produce amplitudes in the coupled channels from where we can extract the pole positions in the complex plane by means of an analytical continuation of these amplitudes. Even then, the problem is subtle because using standard periodic boundary conditions, and a wide range of lattice volumes, there is a gap of energies in the levels of the box precisely for the energies where one finds the poles. Because of this problem one is then forced to use either asymmetric boxes or discretization in the moving frame in order to find eigenvalues of the box in the desired region.  In the present paper we face all these problems and come out with some strategies that we find better suited to  determine the position of the two $\Lambda(1405)$ poles.

\section{Formalism}

 In the chiral unitary approach the scattering matrix in coupled
channels is given by the Bethe-Salpeter equation in its factorized form

\be
T=[1-VG]^{-1}V= [V^{-1}-G]^{-1},
\label{bse}
\ee
where $V$ is the matrix for the transition potentials between the
channels and 
$G$ 
is a diagonal matrix with the $i^{\rm th}$
element, $G_i$,
given by the loop function of two propagators,
a pseudoscalar meson and a baryon,
which is defined as 
\be
\label{loop}
G_i=i 2 M_i\,\int\frac{d^4 p}{(2\pi)^4} \,
\frac{1}{(P-p)^2-M_i^2+i\epsilon}\,\frac{1}{p^2-m_i^2+i\epsilon}
\ ,
%\label{eq:Gl}
\ee
where $m_i$ and $M_i$ are the masses of the meson and the baryon, 
respectively, and $P$ is the four-momentum of the global meson-baryon system.

The loop function in Eq.~(\ref{loop}) needs to be regularized and
this can be accomplished either with dimensional regularization
or with a three-momentum
cutoff. The equivalence
 of both methods was
shown in Refs.~\cite{ollerulf,ramonetiam}.
In dimensional regularization the integral of Eq.~(\ref{loop}) is evaluated and gives for meson-baryon systems
\cite{ollerulf,bennhold}

\begin{align}
\label{eq:g-function}
 &G^D_i(s, m_i, M_i) =  \frac{2M_i}{(4 \pi)^2}
 \Biggr\{a_i(\mu) + \log \frac{m_i^2}{\mu^2} \nn\\ 
&\quad + \frac{M_i^2 - m_i^2 + s}{2s} \log \frac{M_i^2}{m_i^2}\nn \\
 &\quad + \frac{Q_i(\rts)}{\rts}
    \bigg[
         \log \left(  s-(M_i^2-m_i^2) + 2 \rts Q_i(\rts) \right)\nn\\
     &\quad +  \log \left(  s+(M_i^2-m_i^2) + 2 \rts Q_i(\rts) \right)
    \nonumber \\
  &\quad - \log \left( -s+(M_i^2-m_i^2) + 2 \rts Q_i(\rts) \right)\nn\\
&\quad  - \log \left( -s-(M_i^2-m_i^2) + 2 \rts Q_i(\rts) \right)
    \bigg]
  \Biggr\},
\end{align}
where $s=E^2$, with $E$ being the energy of the system in the center of mass
frame, $Q_i$ being the on shell momentum of the particles in the channel $i$,
$\mu$  a regularization scale and $a_i(\mu)$ being a subtraction constant
(note that there is only one degree of freedom, not two independent
parameters).

In other works one uses regularization with a cutoff in three momentum,
once the $p^0$ integration is analytically performed \cite{npa}, and one
gets

\ba
&&G_i=\hspace{-4mm}\int\limits_{|\vec p|<p_{\rm max}}
\frac{d^3\vec p}{(2\pi)^3}\frac{2M_i}{2\omega_1(\vec p)\,\omega_2(\vec p)}
\frac{\omega_1(\vec p)+\omega_2(\vec p)}
{E^2-(\omega_1(\vec p)+\omega_2(\vec p))^2+i\epsilon},
\non 
&&\omega_{1,2}(\vec p)=\sqrt{m_{1,2}^2+\vec p^{\,\,2}}\, ,
\label{prop_cont}
\ea
with $m_1$, $m_2$ corresponding to $m_i$ and $M_i$ of Eq.~(\ref{loop}).

When one wants to obtain the energy levels in the finite box,
 instead of integrating over the
energy states of the continuum, with $p$ being a continuous variable
as in Eq.~(\ref{prop_cont}), one must sum over 
the discrete momenta allowed
in a finite box of side $L$ with periodic boundary conditions.
We then have to replace $G$ by 
$\widetilde G={\rm diag}\,(\widetilde G_1,\widetilde G_2)$ (in two channels), where 
\ba
\widetilde G_{i}&=&\frac{2M_i}{L^3}\sum_{\vec p}^{|\vec p|<p_{\rm max}}
\frac{1}{2\omega_1(\vec p)\,\omega_2(\vec p)}\,\,
\frac{\omega_1(\vec p)+\omega_2(\vec p)}
{E^2-(\omega_1(\vec p)+\omega_2(\vec p))^2},
\non 
\vec p&=&\frac{2\pi}{L}\,\vec n,
\quad\vec n\in \mathds{Z}^3 \,
\label{tildeg}
\ea

 This is the procedure followed in Ref.~\cite{misha}.  The eigenenergies of
the box correspond to energies  that produce poles in the $T$ matrix, 
Eq.~(\ref{bse}), which correspond to zeros of the
determinant of $1-V\widetilde G$,
 \be
\label{eq:det}
\det(1-V\tilde G)=0\, .
\ee  
For the case of two coupled channels Eq.~(\ref{eq:det}) can be written as
\begin{align}
\det(1-V\tilde G)&=1-V_{11}\tilde G_1-V_{22}\tilde G_2\nonumber\\
&\quad+(V_{11}V_{22}-V_{12}^2)\tilde G_1\tilde G_2\nonumber\\
&=0\,.\label{det2}
\end{align} 
The problem of the $\bar K N$ interaction with its coupled channels and the 
$\Lambda(1405)$ was addressed in Ref.~\cite{angelskaon} using the cut off method, but more recently it has been addressed using dimensional regularization \cite{ollerulf,bennhold}. For this reason we will also use the dimensional regularization method for the finite box, which was developed in Ref.~\cite{alberto}. The change to be made is also very simple, the $G$ function of dimensional regularization of Eq. (\ref{eq:g-function}) has to be substituted by 

\begin{align}
\widetilde G(E)&=G^D(E)+\lim_{p_\textrm{max}\to \infty}
\Bigg[\frac{1}{L^3}\sum_{\vec{p}_i}^{p_\textrm{max}}I(p_i)\nonumber\\
&\quad-\int\limits_{p<p_\textrm{max}}\frac{d^3p}{(2\pi)^3} I(p)\Bigg]
\label{tonediff}
\end{align}

where $I(p)$ is given by 
 \be
I(p)=\frac{2M_i}{2\omega_1(\vec p)\,\omega_2(\vec p)}
\frac{\omega_1(\vec p)+\omega_2(\vec p)}
{E^2-(\omega_1(\vec p)+\omega_2(\vec p))^2+i\epsilon}.
\label{prop_contado}
\ee

We will also consider the case where the meson-baryon system moves with a fourmomentum $P=(P^0, \vec P)$ in the box. In this case we still have to define the integrals and the sums in the CM frame, where $p_{max}$ is defined, but the momenta of the two particles must be discretized in the box, where the system moves with momentum $P$. We follow the approach of Refs.~\cite{mishamoving,luisroca} and use the boost transformation from the moving frame, with the variables $\vec{p}_{1,2}$, to the CM frame with the variables $\vec{p}^{\,*}_{1,2}$

\be
\vec{p}^{\,*}_{1,2}=\vec p_{1,2} + \left[\left(\frac{M_I}{P^0}-1\right)
\frac{\vec p_{1,2}\cdot\vec P}{|\vec P|^2}
-\frac{p_{1,2}^{*0}}{P^0}\right]\vec P.
\label{boostmisha}
\ee
where $M_I^2=P^2={P^0}^2-\vec P\,^2$, the subindexes
$1,2,$ represent the meson, baryon particles and  $p^{*0}_{1,2}$ are the CM energies of the particles given by
\be
p^{*0}_{1,2}=\frac{M_I^2+m_{1,2}^2-m_{2,1}^2}{2M_I}.
\label{q0cm}
\ee
Then we must do the substitution in Eq. (\ref{tonediff})
for the evaluation of the energies in the box,

\begin{align}
 \lim_{p_\textrm{max}\to\infty} \frac{1}{L^3}\sum_{\vec{p}_i}^{p_\textrm{max}}I(p_i)
\longrightarrow &\frac{1}{L^3}\sum_{\vec p}^{|\vec {p}^{\,*}|<p_{\rm max}}\frac{ M_I}{P^0}I(p^*_i),%\quad 
\non
&\vec p=\frac{2\pi}{L}\vec n,\quad \vec n\in \mathds{Z}^3\label{smallp}
\end{align}
with $\vec{p}^{\,*}_i$ given in terms of $\vec{p}_i$ by means of Eq. (\ref{boostmisha}).

Since $\vec{p}_1$ and $\vec{p}_2=\vec{P}-\vec{p}_1$ must both satisfy the periodic boundary conditions, this forces $\vec P$ to be also discretized and thus we can only use values of $\vec P$ such that 

\be
\vec P=\frac{2\pi}{L}\vec N,\quad \vec N\in \mathds{Z}^3~.\label{totalP}
\ee

\section{Results}
\subsection{Energy levels in the box}
In this section we show the energy levels obtained from the solution of
Eq.~(\ref{eq:det}) as a function of the side length of the box, $L$, and for different physical cases: using periodic boundary conditions in a: (1) symmetric box, (2) asymmetric box and (3) symmetric box but in a moving frame, i.e., with non-zero value for the total center of mass momentum $\vec{P}$ (Eq.~(\ref{totalP})).

\subsubsection{periodic boundary conditions in a symmetric  box}
\begin{figure}
\includegraphics[width=0.5\textwidth]{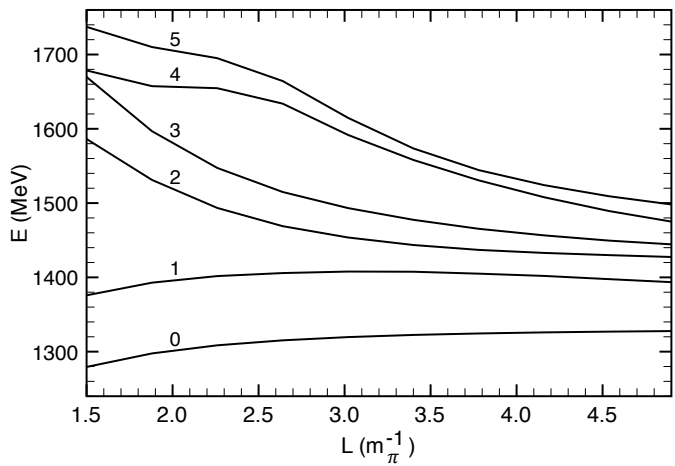}
\caption{Energy levels in a symmetric box of side length $L$.}\label{chiral_levels_symmetric}
\end{figure}

In Fig.~\ref{chiral_levels_symmetric} we show the first six energy levels related to the system formed by the coupled channels $\bar K N$, $\pi\Sigma$, $\eta\Lambda$ and $K\Xi$, which generate a double pole structure for the $\Lambda(1405)$ and a pole for the $\Lambda(1670)$ \cite{cola}. These levels are obtained by solving Eq.~(\ref{eq:det}) using the chiral model of Ref.~\cite{cola} and imposing periodic boundary conditions in a symmetric  box of side length $L$ (measured in units of $m^{-1}_\pi$). 

As can be seen in Fig.~\ref{chiral_levels_symmetric}, the gap between the levels 0, 1 and especially between levels 1 and 2 is considerable, giving rise to the presence of only two levels in the energy region of interest, i.e., the energy range in which the two poles of the $\Lambda(1405)$ are found ($1390-1430$ MeV).  This fact shows the difficulty that one can face to extract information about the poles of the $\Lambda(1405)$ in an infinite volume considering these energy levels as reference.

\subsubsection{periodic boundary conditions in an asymmetric  box}
\begin{figure}
\includegraphics[width=0.5\textwidth]{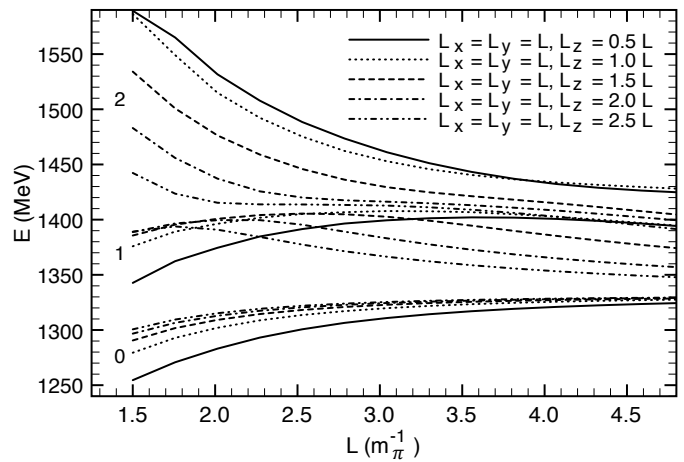}
\caption{Energy levels in an asymmetric box of side length $L_x=L_y=L$ and $L_z=zL$, with $z=0.5-2.5$ in steps of $0.5$.}\label{chiral_levels_asymmetric}
\end{figure}

To see if we can obtain more energy levels in the region of the $\Lambda(1405)$, it is also possible to solve Eq.~(\ref{eq:det}) but in an asymmetric box. To do this we just need to substitute $L^3$ by $L_xL_yL_z$ and the momentum $\vec{p}$ of Eq.~(\ref{smallp}) by $\vec{p}=(2\pi)(n_x/L_x,n_y/L_y,n_z/L_z)$. In Fig.~\ref{chiral_levels_asymmetric} we show the first three energy levels determined in a box of side lengths $L_x=L_y=L$ and $L_z=zL$, and we vary $z$ between $0.5L$ and $2.5L$. In this way, we get more energy levels in the region of interest, which can provide different information about the system and the poles of the $\Lambda(1405)$.

\subsubsection{periodic boundary conditions in a moving frame}
\begin{figure}
\includegraphics[width=0.5\textwidth]{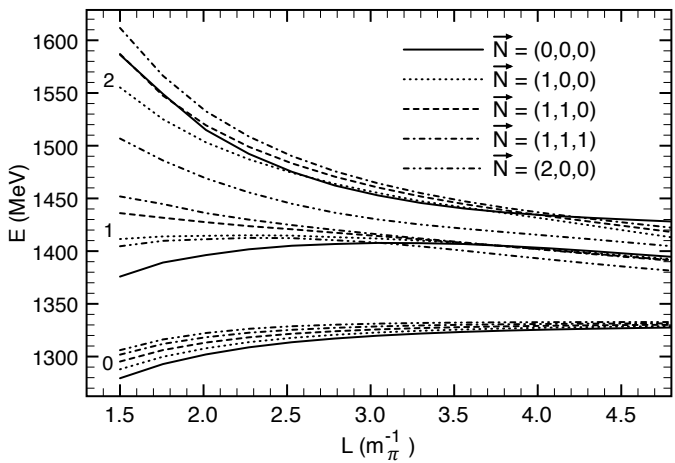}
\caption{Energy levels in a symmetric box of side length $L$ with the system having a center of mass momentum given by Eq.~(\ref{totalP}).}\label{chiral_levels_movingframe}
\end{figure}

Another method to try to get more energy levels around the pole positions of the $\Lambda(1405)$ and thus, different information about the dynamics of the system under consideration, consists of imposing periodic boundary conditions in a symmetric box of side length $L$ but considering the system in a moving frame, i.e., with non zero center of mass momentum $\vec{P}$. In Fig.~\ref{chiral_levels_movingframe} we show the results found in this case for the first three levels obtained and for different values of the vector $\vec{N}$ (see Eq.~(\ref{totalP})). As can be seen,  the use of different values of $\vec{P}$ gives rise to a splitting of the levels. In particular, the splitting of level 1 is precisely in the energy region of interest, $1390-1450$ MeV.  This is different from the case of the asymmetric box, where level 2 is required in order to have energy levels around 1420-1450 MeV, as can be seen in Fig.~\ref{chiral_levels_asymmetric}. 
\subsection{The inverse problem: getting the $\Lambda(1405)$ poles from the energy levels of the box}
In the following we refer to the problem of determining the pole positions of the $\Lambda(1405)$ in the infinite volume using the energy levels shown in Figs.~\ref{chiral_levels_symmetric},~\ref{chiral_levels_asymmetric},~\ref{chiral_levels_movingframe} as if they were provided to us by a lattice calculation. In our formalism, we can simulate lattice-like data considering points related to the energy levels of Figs.~\ref{chiral_levels_symmetric},~\ref{chiral_levels_asymmetric},~\ref{chiral_levels_movingframe} and assigning to them a typical error of $\pm$ 10 MeV.  We call the data generated in this form ``synthetic" lattice data and the problem of getting the poles of the $\Lambda(1405)$ from these data points  ``the inverse problem". 

To solve the inverse problem we consider a potential with the same energy dependence than the chiral potential used to generate the energy levels shown in Figs.~\ref{chiral_levels_symmetric},~\ref{chiral_levels_asymmetric},~\ref{chiral_levels_movingframe}. In its non-relativistic version, this potential is given by \cite{angelskaon}

\begin{equation}
V_{ij}=-\frac{C_{ij}}{4f^2}(E_i+E_j),\label{Vchiral}
\end{equation}
with  $C_{ij}$ coefficients depending on the channel considered, $f$ being the pion decay constant and $E_i$ ($E_j$)  being the center of mass energy of the meson in the initial (final) state. Using that for a particular channel $l$
\begin{align}
E_l&=\frac{E^2+m^2_l-M^2_l}{2E},\nonumber\\
\end{align}
with $m_l$ and $M_l$ being the masses of the meson and baryon which constitute channel $l$, respectively, we can write Eq.~(\ref{Vchiral}) as

\begin{equation}
V_{ij}=-\frac{C_{ij}}{4f^2}\left\{E+\frac{1}{2E}\left[m^2_i+m^2_j-(M^2_i+M^2_j)\right]\right\}\label{VE}
\end{equation}
Choosing a region of energies around a certain value of $E$, $E_0$, the inverse function of $E$ can be expanded as a function of $E-E_0$ to a good extent. Particularizing $E_0$ to the value given by the sum of the kaon and nucleon masses, i.e., $m_K+M_N$, we can  write the potential in Eq.~(\ref{VE}) as

\begin{equation}
V_{ij}=a_{ij}+b_{ij}[E-(m_K+M_N)].\label{Vpara}
\end{equation}
The value of the coefficients $a_{ij}$ and $b_{ij}$ can be obtained comparing Eq.~(\ref{Vpara}) with Eq.~(\ref{VE}) and substituting $1/E$ by its Taylor expansion around
$E_0=m_K+M_N$.

To solve the inverse problem, we use the energy levels obtained from Eq.~(\ref{eq:det}) with the potential of Eq.~(\ref{Vpara}) but treat $a_{ij}$ and $b_{ij}$ as parameters which are determined by fitting the corresponding solutions for the energy levels to the ``synthetic" lattice data considered. Since this potential has the same energy dependence as the chiral potential, the best fit we can perform will have as minimum value for the $\chi^2$ the result $\chi^2_{min}=0$. However, other possible potentials, giving rise to solutions compatible with the error assumed in the data points, can be also found as an answer for the inverse problem. These solutions can be obtained by generating random numbers for the parameters $a_{ij}$ and $b_{ij}$ close to those of the minimum such that $\chi^2\leqslant\chi^2_{min}+1$.

It is important to notice that the loop function $\tilde{G}$, used in Eq.~(\ref{eq:det}), needs to be regularized and, thus, depends on a cut-off or a subtraction constant. Consequently, so do the  fitted parameters, but the $T$ matrix obtained from Eq.~(\ref{bse})  and the observables related to it should be independent of this regularization parameter.  This means that the inverse method cannot depend on the cut-off or subtraction constant assumed in the evaluation of  the $\tilde{G}$ function.  For the case of one channel, it is possible to show analytically this independence in the choice of the cut-off or subtraction constant \cite{misha,alberto}, but if more channels are involved it can only be seen numerically
by changing the cut-off or subtraction constant in a reasonable physical range \cite{misha,alberto}. 

In the next sections we show the results found for the inverse problem. To accomplish this we have considered different sets of points extracted from the energy levels shown in Figs.~\ref{chiral_levels_symmetric},~\ref{chiral_levels_asymmetric},~\ref{chiral_levels_movingframe} and fit them from the solution that Eq.~(\ref{eq:det}) produces with the potential  of Eq.~(\ref{Vpara}). To solve Eq.~(\ref{eq:det}) we have taken into account two coupled channels, $\pi\Sigma$ (which we named channel 1) and $\bar K N$ (or channel 2), which are the most relevant channels to describe the properties of the $\Lambda(1405)$.  This implies, as can be seen in Eq.~(\ref{det2}), that we have to determine three potentials, $V_{11}$, $V_{12}$ ($V_{21}=V_{12}$) and $V_{22}$ or equivalently 6 parameters $a_{11}$, $a_{12}$, $a_{22}$, $b_{11}$, $b_{12}$ and $b_{22}$. Once the parameters and, thus, the potentials, are known, we can use them to solve Eq.~(\ref{bse}) and determine the pole positions of the $\Lambda(1405)$ in an infinite volume.
\subsubsection{periodic boundary conditions in a symmetric  box}
\begin{figure}
\includegraphics[width=0.5\textwidth]{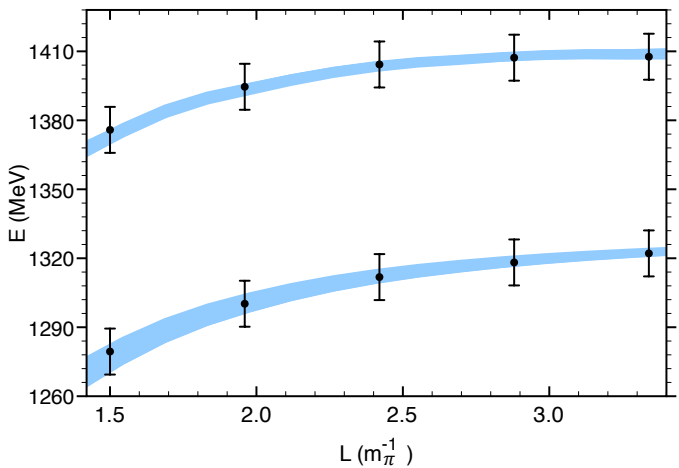}
\caption{(Color online) First two energy levels as function of the box side length $L$, reconstructed from fits to the ``synthetic" data of Fig.~\ref{chiral_levels_symmetric} in a range of $L$ between $1.5\,m^{-1}_{\pi}$ and $3.34\,m^{-1}_{\pi}$ using the potential of Eq.~(\ref{Vpara}). The band corresponds to different choices of parameters within errors.}\label{ressym1}
\end{figure}

In Fig.~\ref{ressym1} we show the results of the energy levels reconstructed from the best fits to the ``synthetic" lattice data considered from Fig.~\ref{chiral_levels_symmetric}. These data consist of 10 points for levels 0 and 1 obtained in a symmetric box of side length $L$, varying $L$ in the range $1.5\,m^{-1}_{\pi}$ to $3.34\,m^{-1}_{\pi}$, assigning an error of $\pm~ 10$ MeV to the eigenenergies of the box (from now on, we will always assume an error of $\pm~ 10$ for the different ``synthetic" data that we will use). The shadowed band in the figure corresponds to the random choices of parameters satisfying the condition $\chi^2\leqslant\chi^2_{min}+1$.
Using the potentials obtained from the fit and the loop function $G$ in infinite volume, we can solve Eq.~(\ref{bse}) and calculate the two-body $T$ matrix in the unphysical sheet, which allows us to determine the pole position of the $\Lambda(1405)$ associated to the band of solutions shown in Fig.~\ref{ressym1}. As a result we get a double pole structure for the $\Lambda(1405)$, with one pole in the region 1385-1433 MeV and half width between 93-137 MeV (which we call pole 1) and another one in the energy region 1416-1427 MeV and half width in the range 11-20 MeV (which we call pole 2). If we compare these results with the ones of the chiral model~\cite{cola}, $1390-i66$ MeV and $1426-i16$ MeV, respectively, we find a big dispersion in the determination of the real part of the first pole of the $\Lambda(1405)$. This shows that the information which one can extract from the ``synthetic" data considered in Fig.~\ref{ressym1} is not sufficient to determine with more precision the poles associated with the $\Lambda(1405)$. 

\begin{figure}
\includegraphics[width=0.5\textwidth]{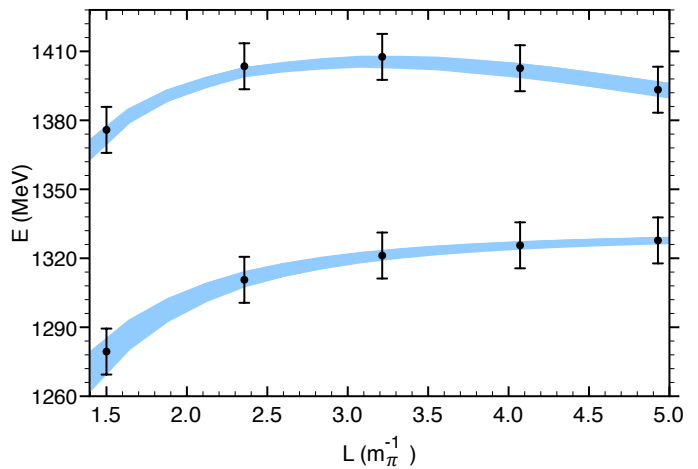}
\caption{(Color online) Same as in Fig.~\ref{ressym1} but for a range of $L$ between $1.5\,m^{-1}_{\pi}$ and $4.93\,m^{-1}_{\pi}$.}\label{ressym2}
\end{figure}

A way to delimit the poles of the $\Lambda(1405)$ with more precision from lattice data could consist in going to higher volumes, since for large volumes
the results in the box should be very close to those of an infinite volume. With this idea in mind, we can generate ``synthetic" data points for the levels 0 and 1 of Fig.~\ref{chiral_levels_symmetric}, but in a larger range of $L$ than that considered in Fig.~\ref{ressym1}. The data points, as well as the results from the fits, are shown in Fig.~\ref{ressym2}.
Similarly, if we use now the potentials associated with the band of solutions shown in Fig.~\ref{ressym2} to solve Eq.~(\ref{bse}) and calculate the $T$ matrix in the unphysical sheet, we get again two poles in the complex energy plane associated with the $\Lambda(1405)$: one in the region 1390-1433 MeV, with half width between 70-100 MeV, and other at 1410-1421 MeV with half width 17-30 MeV. Comparing them with the previous results, we find that the consideration of a bigger box has improved slightly the width associated with the first pole of the $\Lambda(1405)$, however, we continue having a similar energy dispersion for the real part of the pole.

\begin{figure}
\includegraphics[width=0.5\textwidth]{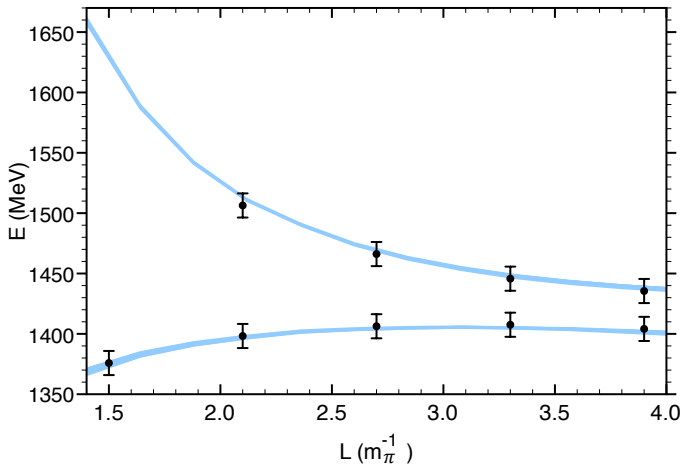}
\caption{(Color online) Fits to the levels 1 and 2 of Fig.~\ref{chiral_levels_symmetric} constructed from the potential of Eq.~(\ref{Vpara}).}\label{ressym3}
\end{figure}

We could also try using different levels than those employed in Figs.~\ref{ressym1} and \ref{ressym2} to see if we can get more reliable information from them.  In Fig.~\ref{ressym3} we consider  ``synthetic" data obtained from levels 1 and 2 of Fig.~\ref{chiral_levels_symmetric}. We have taken into account 5 points for level 1 in a range of $L$ between $1.5\,m^{-1}_{\pi}$ and $3.9\,m^{-1}_{\pi}$ and 4 points for level 2 for values of $L$ inside  $2\,m^{-1}_{\pi}$ to $3.9\,m^{-1}_{\pi}$. This is because for level 2 the points for values of $L$ below $2\,m^{-1}_{\pi}$ are influenced by the $\eta\Lambda$ and $K\Xi$ channels and, thus, it is not possible to fit them considering only the $\pi\Sigma$ and $\bar K N$ channels, as we do. We can use now the potentials associated with the different fits shown in Fig.~\ref{ressym3} to calculate the pole positions of the $\Lambda(1405)$ in infinite volume by means of Eq.~(\ref{bse}). In this case, we continue getting a double pole structure for the $\Lambda(1405)$, but this time one pole is at $(1375-1430) - i (70-85)$ MeV and the other one is at $(1412-1427)-i(21-34)$ MeV.  The position of the second pole remains basically the same as in the two previous cases. However, the use of ``synthetic" points generated from levels 1 and 2 instead than from levels 0 and 1 has restricted more the imaginary part of the first pole, although we continue getting a similar energy dispersion for the real part of the pole position. 

\begin{figure}
\includegraphics[width=0.5\textwidth]{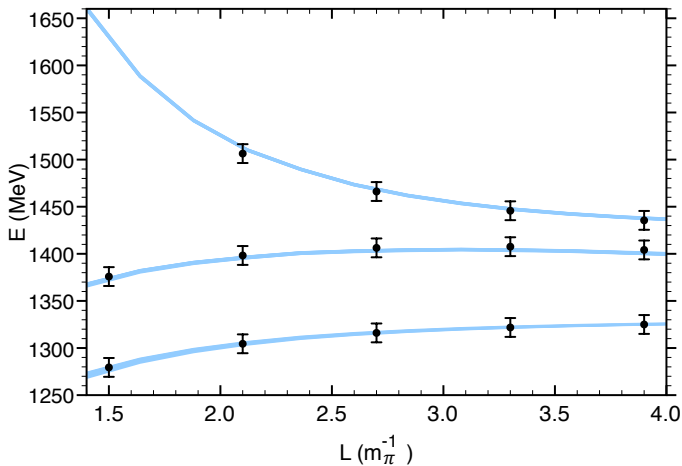}
\caption{(Color online) Fits to the levels 0, 1 and 2 of Fig.~\ref{chiral_levels_symmetric} constructed from the potential of Eq.~(\ref{Vpara}).}\label{ressym4}
\end{figure}

Finally, we could consider all the energy levels present in Fig.~\ref{chiral_levels_symmetric} below 1600 MeV to generate data points to check if the consideration of more levels can restrict more the energy region at which the pole positions of the $\Lambda(1405)$ are found. Following this idea, in Fig.~\ref{ressym4} we consider a set of 14 points extracted from levels 0, 1 and 2 of Fig.~\ref{chiral_levels_symmetric}. Similar to the previous results, the consideration of data points associated to three energy levels puts a restriction on the imaginary part of the first pole of the $\Lambda(1405)$, which in this case is in the range 54-68 MeV (closer to the chiral solution, 66 MeV). However the dispersion on the real part continues basically equal, 1400-1428 MeV. For the second pole we get ($1408-1425)-i(29-40)$ MeV.

These results show that the information which can be extracted from ``synthetic" data constructed from the energy levels obtained in a symmetric box of volume $L^3$ is not enough to determine with precision the pole positions of the $\Lambda(1405)$, a  fact which is basically related to the absence of energy levels, and thus information about the dynamics of the system, in the region between 1400-1500 MeV, as can be seen in Fig.~\ref{chiral_levels_symmetric}.
  
\subsubsection{periodic boundary conditions in an asymmetric  box}
\begin{figure}
\includegraphics[width=0.5\textwidth]{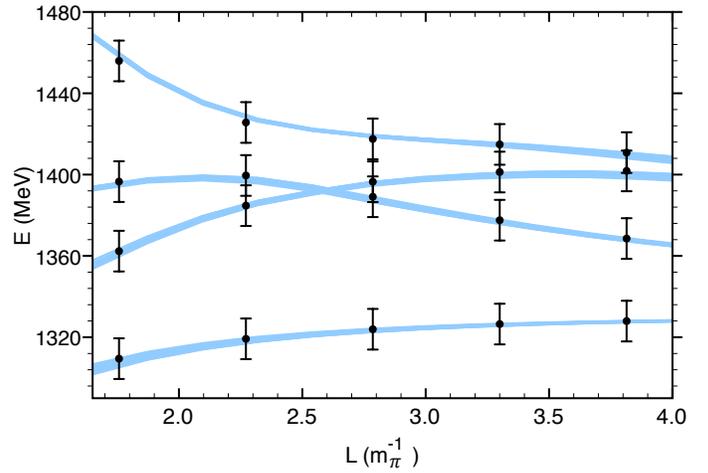}
\caption{(Color online) Fits to the ``synthetic" data extracted from the energy levels 0, 1 and 2 of Fig.~\ref{chiral_levels_asymmetric} in an asymmetric box of side lengths $L_x=L_y=L$ and $L_z=zL$. The data points considered are generated from level 0 for $z=2.5$, level 1 for $z=0.5$ and $z=2.0$ and level 2 for $z=2.0$.}\label{resasym}
\end{figure}

We consider now the case of an asymmetric box of side lengths $L_x=L_y=L$ and $L_z=zL$ to solve the inverse problem. In this case, we generate a set of 20 data points extracted from levels 0, 1 and 2 shown in Fig.~\ref{chiral_levels_asymmetric}. In particular, we use 5 points for level 0 calculated with $z=2.5$, 10 points for level 1 (5 for the case $z=0.5$ and 5 more for $z=2.0$) and 5 points for level 2 obtained with $z=2.0$. In this way we ensure the presence of some energy levels in the region of the $\Lambda(1405)$, as can be seen in Fig.~\ref{resasym}.

The solution of the Bethe-Salpeter equation in an infinite volume, Eq.~(\ref{bse}), using the potentials related to the band of solutions plotted in Fig.~\ref{resasym} shows the presence of a double pole structure for the $\Lambda(1405)$ with pole positions at $(1383-1407) -i (57-69)$ MeV and $(1425-1434)-i (25-35)$ MeV. Thus, using this new set of data points, there is an improvement in the determination of the first pole of the $\Lambda(1405)$, which is now quite close to the chiral result ($1390-i66$ MeV). However, the second pole appears at higher energies as compared to the case of a symmetric box and sometimes is far from the chiral solution ($1426-i16$ MeV), being even close to the $\bar K N$ threshold.

\subsubsection{periodic boundary conditions in a moving frame}
\begin{figure}
\includegraphics[width=0.5\textwidth]{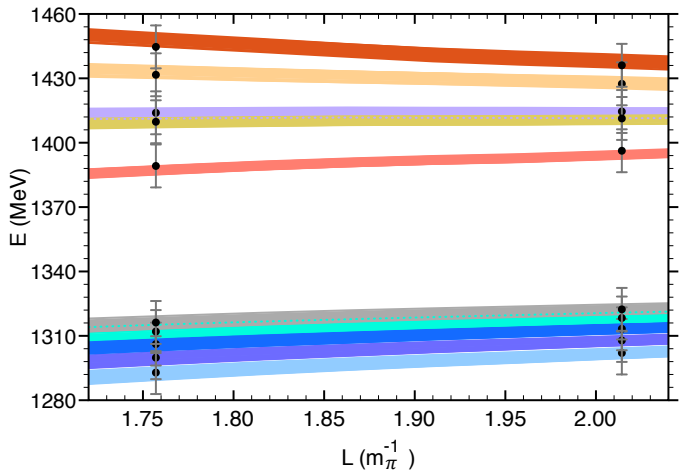}
\caption{(Color online) Fits to the ``synthetic" data extracted from the energy levels 0 and 1 of Fig.~\ref{chiral_levels_movingframe}, which correspond to the case of a symmetric box, but with the particles being in a moving frame.}\label{Fitmoving}
\end{figure}

We can also study the information which can be extracted for the poles of the $\Lambda(1405)$ using the levels obtained when we consider the system in a symmetric box, but in a moving frame, to generate ``synthetic" lattice data. In this case, we consider levels 0 and 1 of Fig.~\ref{chiral_levels_movingframe} determined for 5 different values of the center of mass momentum (the ones shown in the legend of Fig.~\ref{chiral_levels_movingframe}) and two points in each of these curves. In particular, we take points at $L=1.757 \,m^{-1}_{\pi}$ and $L=2.014\,m^{-1}_{\pi}$, obtaining then 20 data points. The results are shown in Fig.~\ref{Fitmoving}. From the solution of the best fits, we can use the potentials obtained to solve Eq.~(\ref{bse}), getting then two poles for the $\Lambda(1405)$: one at $(1388-1418) -i(59-77)$ MeV and other at $(1412-1427)-i(16-34)$ MeV.

\begin{figure}
\includegraphics[width=0.5\textwidth]{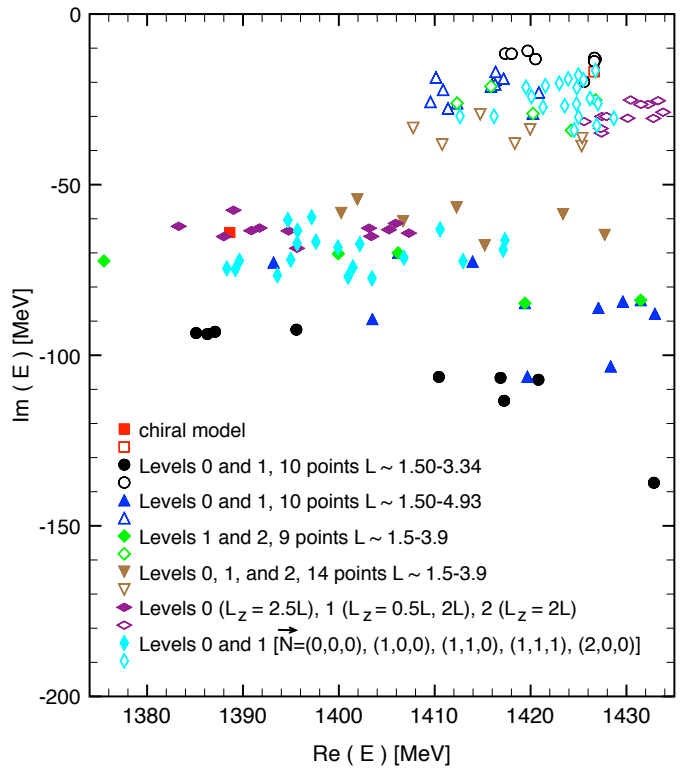}
\caption{(Color online) Pole positions of the $\Lambda(1405)$ reconstructed from the different set of ``synthetic" data generated for the different cases considered in this work. The shaded symbols corresponds to the positions obtained for the first pole of the $\Lambda(1405)$, while the empty symbols are related to the second pole of the $\Lambda(1405)$.}\label{poles}
\end{figure}

In Fig.~\ref{poles} we show the results for the pole positions of the $\Lambda(1405)$ obtained from the different data set considered in this work.
As can be seen in Fig.~\ref{poles}, out of the different data sets considered to solve the inverse problem, the cases of an asymmetric box and of a symmetric box but in a moving frame seem to be more suited to get the two poles of the $\Lambda(1405)$ with more precision.

\subsection{Phase shifts and L\"uscher approach}
\begin{figure}
\includegraphics[width=0.45\textwidth]{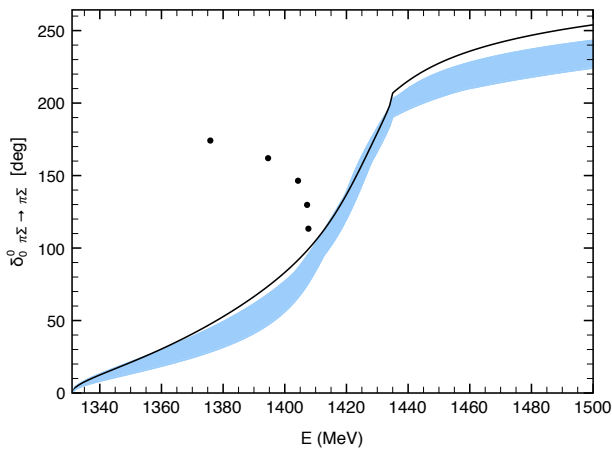}
\caption{(Color online) s-wave phase shift for the $\pi\Sigma$ channel in isospin 0 determined from the chiral model (solid line), using the band of solutions for the energy levels shown in Fig.~\ref{ressym1} (shaded area) and within the L\"uscher formula (dots).}\label{phasesym}
\end{figure}

So far,  when solving the inverse problem, we have determined the pole positions related to the $\Lambda(1405)$ using the two-body $T$ matrix in infinite volume obtained from the energy levels calculated in a finite volume. However, we could also make use of the same scattering matrix to obtain the phase shifts in infinite volume for the different channels. To do this, we follow Ref.~\cite{npa} in which the $T$ matrix is related to the inelasticity and the phase shifts\footnote{ Please note that in Ref.~\cite{npa} coupled meson-meson systems are studied while, in the present case, we analyzed coupled meson-baryon systems. Thus, a different normalization shall be used in the determination of the phase-shifts: instead of the scattering matrix  $T_{ij}$ for the transition between the channels $i$ and $j$, a factor $\sqrt{2 M_i 2M_j}$ shall be included (with $M_i$, $M_j$ being the baryon masses in the channels $i$, $j$), thus $T_{ij}\to \sqrt{2 M_i 2M_j} T_{ij} $.}. Just as an illustrative example, we can use the two coupled channel  $T$ matrices obtained from the band of solutions shown in Fig.~\ref{ressym1} and calculate the phase shift, for example, for the $\pi\Sigma$ channel. In Fig.~\ref{phasesym} we show the phase shift for the $\pi\Sigma\to\pi\Sigma$ transition in isospin 0 and s-wave, $\delta^0_0$, in the energy region 1331 to 1500 MeV, determined within the chiral model (solid line) and using the inverse method from the band of energies shown in Fig.~\ref{ressym1} (shaded area). As can be seen, the agreement between the theoretical curve and the solutions determined using the fits to the ``synthetic" data of Fig.~\ref{ressym1} is good. This is more remarkable having in mind the fact that the data points considered in Fig.~\ref{ressym1} reach up to a maximum energy of around 1410 MeV. To determine the $\pi\Sigma$ phase shift
with more precision from eigenenergies obtained in a finite volume more energy levels at higher energies than 1410 MeV would be required. However, the information contained in the ``synthetic" data of Fig.~\ref{ressym1} about the dynamics of the system makes possible to get an overall agreement with the theoretical $\pi\Sigma$ phase shift for energies bigger than the one of the last data point considered in Fig.~\ref{ressym1}.

One may wonder whether the use of the L\"uscher formula for the single $\pi\Sigma$ channel can give rise to similar phase shifts as those shown in Fig.~\ref{phasesym} (solid line or shaded area).  As we have seen, the poles of the $\Lambda(1405)$ come as a consequence of the $\pi\Sigma$ and $\bar K N$ coupled channel dynamics, but we have seen in the analogous case of the s-wave $I=0$ $\pi \pi$ scattering, that the $\sigma(600)$ and $f_0(980)$ resonances also require the analysis with the $\pi \pi$ and $K \bar K$ channels, but at low energies the $\pi \pi$ phase shifts are very accurately obtained from finite volume energy levels via the L\"uscher formula with just the $\pi \pi$ channel \cite{misha}. We might think that this is the case here too, and it is generally accepted that the L\"uscher formula can be used when one has only one channel open. Assuming that this is the case, we proceed as follows: As shown in Ref.~\cite{misha}, for a one channel case and for the discrete eigenenergies obtained in a finite volume, the scattering matrix in infinite volume can be written as

\begin{equation}
T(E)=[\tilde{G}(E)-G(E)]^{-1}.\label{tluescher}
\end{equation}

As proved in Ref.~\cite{misha}, Eq. (\ref{tluescher}) is equivalent to L\"uscher formula. Using now Eq.~(\ref{tluescher}), we can calculate the phase shift for the $\pi\Sigma$ channel for each of the eigenenergies (``synthetic" data points) shown in Fig.~\ref{ressym1} which fall above the $\pi\Sigma$ threshold. This means only the upper level since the energies of the lower level are below the $\pi\Sigma$ threshold.
The results obtained are shown as dots in Fig.~\ref{phasesym}. As can be seen, the phase shift determined within the L\"uscher formula, Eq.~(\ref{tluescher}), have a completely different behavior than the theoretical result determined from the chiral model (solid line) or those of the shaded area calculated from the fits to  the ``synthetic" lattice data of Fig.~\ref{ressym1}. This fact indicates that the $\pi\Sigma$ and $\bar K N$ coupled channel dynamics plays a very important role in the generation of the $\Lambda(1405)$ and, thus, in the determination of physical observables like the phase shifts.

   The reason for this failure is that the higher level of Fig.~\ref{ressym1} is mostly tied to the $\bar K N$ channel and, hence, forcing it to provide information on $\pi\Sigma$ leads to unrealistic results. This is an important result which could not be anticipated and tells us that the straightforward application of L\"uscher formula to obtain phase shifts can provide quite unrealistic results if applied in physical cases where two or more channels are very much connected. The value of the present approach to make prospective studies in different cases becomes apparent. However, the discussion requires a closer look, as we argue below.\\

\section{Further discussions}
The results obtained  in the previous subsection could be surprising to the light of a well known fact in coupled channels 
studies where the effect of coupled channels that one would like to disregard in the analysis can be reabsorbed by a redefinition of the potential in a chosen channel. Imagine we have $N$ channels and we want to include the effect of $(N-1)$ channels redefining the potential of channel 1. This is done in Refs.~\cite{HyodoW,BayarX} and the effective potential for channel 1 is now given by

\begin{align}
V_{\textrm{eff}}&=V_{11}+\sum_{m=2}^N V_{1\,m}G_m V_{m\,1}\nonumber\\
&\quad+\sum_{m,\,l=2}^N V_{1\,m}G_m t^{(N-1)}_{m \,l}G_l V_{l\,1},
\end{align}
where $t^{(N-1)}_{m \,l}=[{V^{(N-1)}}^{-1}-G^{(N-1)}]^{-1}$ is the $(N-1)(N-1)$ $t$ matrix of the system of $(N-1)$ channels, after removing channel 1.
This means that it is possible to construct an effective potential and work with just one channel, but it involves $G_m$ and $t^{(N-1)}_{m \,l}$ for the channels not considered.
This has obvious repercussions since for finite volume $G_m$ and $t^{(N-1)}_{m \,l}$ would become $\tilde{G}_m$ and $\tilde{t}^{(N-1)}_{m \,l}$, respectively, which are volume dependent. Then L\"uscher approach is bound to have problems since it implicitly relies upon volume independent potentials.

For the case of finite volume studies it is more practical to state this fact in a different way. Let us start from Eq.~(\ref{bse}) that gives the $T$ matrix in the infinite volume and write the correspondent  scattering matrix in the finite volume, $\tilde{T}$:

\begin{align}
\tilde{T}=[V^{-1}-\tilde{G}]^{-1}. \label{a}
\end{align}
Using Eqs.~(\ref{bse}) and (\ref{a}), we get
\begin{align}
\tilde{T}^{-1}&=T^{-1}-\delta G=T^{-1}[1-T\,\delta G], \label{b}
\end{align}
where we have defined $\delta G\equiv \tilde{G}-G$. Hence,
\begin{align}
\tilde{T}=[1-T\,\delta G]^{-1} T. \label{c}
\end{align}
One can note that this formula is like Eq.~(\ref{bse}), or Eq.~(\ref{a}) for $\tilde{T}$, substituting $V\to T$ and $\tilde{G}\to \delta G$. Hence,
the condition to obtain the energy levels in the box, $\textrm{det}(\tilde{T})=0$, leads to the analogous secular equation of Eq.~(\ref{det2}) in terms of $T$ and $\delta G$ substituting $V$ and $\tilde{G}$, respectively,
\begin{align}
(1-T_{11}\,\delta G_{11})(1-T_{22} \delta G_2)-T^2_{12}\delta G_1 \delta G_2=0,
\end{align}
from where we derive for the eigenenergies of the eigenstates of the box\footnote{We are indebted to the referee for providing us this formula.}
\begin{align}
T_{11}=\delta G_1^{-1}-\frac{T^2_{12}\,\delta G_2}{1-T_{22}\delta G_2}.\label{e}
\end{align}
It is now interesting to note that if channel 2 is a closed channel, like $\bar K N$ below threshold, then $\delta G_2$ is volume exponentially suppressed and in the large $L$ limit we recover L\"uscher formula $T_{11}=\delta G_1^{-1}$.

It is also interesting to establish the connection of $\delta G$ with another function used in finite volume studies, $F^{(d)}(k L)$. Indeed $\delta G$ is proportional to the function $F^{(d)}(k L)$ used in Ref.~\cite{arXiv:1108.5371}, where it is shown in detail that it is actually exponentially suppressed. The superindex $d$ in $F^{(d)}(k L)$ stands for the total momentum (in units of $2\pi/L$) of the two particle system. In Ref.~\cite{arXiv:1108.5371} it is also shown that the use of moving frames and some particular combinations of them can help reducing the volume dependence in the determination of binding energies of systems.

For what respects us in the present case, we can already appreciate in Fig.~\ref{phasesym} that for higher energies, which are reached in Fig.~\ref{ressym1} for large volumes, the $\pi\Sigma$ phase shift determined with L\"uscher formula starts converging to the exact one. However, as can be seen in Eq.~(\ref{e}), the divergence of $T_{11}$ (where, in this case, $1$ represents the $\pi\Sigma$ channel)  
from the L\"uscher formula is tied not only to the exponentially suppressed magnitude $\delta G_2$, but also to the magnitude of $T_{12}$ (transition matrix between $\pi\Sigma$
and $\bar K N$), and thus the actual accuracy of L\"uscher formula to derive the $\pi\Sigma$ phase shift cannot be determined a priori without knowledge of the underlying dynamics.

In view of this, let us investigate this issue further: if the second term in Eq.~(\ref{e}) is suppressed at large $L$ as compared to the first one, the L\"uscher term, how relevant is it then in the determination of the scattering matrix for the different cases considered here: symmetric box, asymmetric box and the system in a moving frame? Can we neglect it for some of these cases and work just with the L\"uscher formula to obtain the scattering matrix?

\begin{figure}
\includegraphics[width=0.45\textwidth]{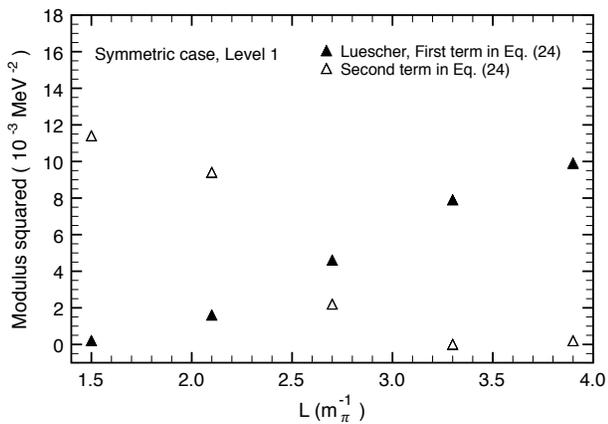}
\caption{Modulus squared of the contributions arising from the first (filled triangles) and second (empty triangles) term of Eq.~(\ref{e}) for the eigenenergies related to the level 1 of Fig.~\ref{ressym4} (symmetric box) as a function of the side length of the box, $L$.}\label{sym24L}
\end{figure}

\begin{figure}
\includegraphics[width=0.45\textwidth]{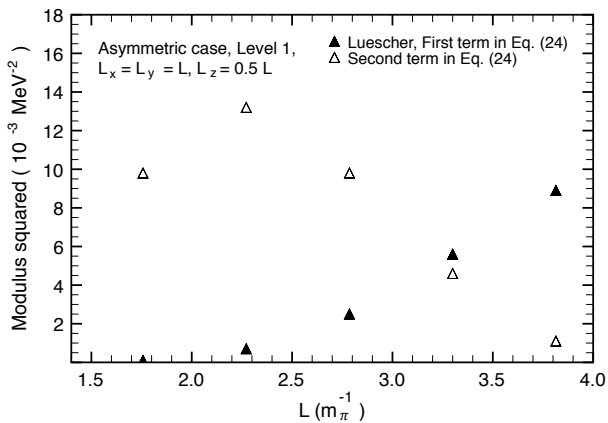}
\caption{Same as in Fig.~\ref{sym24L}, but for the level 1 of an asymmetric box of dimension $L_x=L$, $L_y=L$ and $L_z=0.5 L$ (see Fig.~\ref{resasym}).}\label{asym24L}
\end{figure}

\begin{figure}
\includegraphics[width=0.45\textwidth]{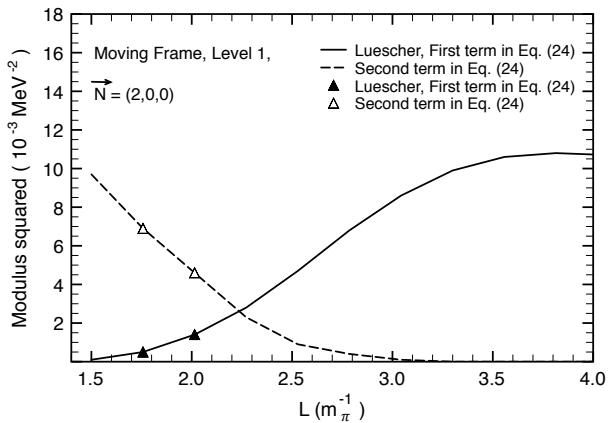}
\caption{Same as in Fig.~\ref{sym24L} but for the level 1 of a system in a moving frame with $\vec{P}=(2\pi/L)(2,0,0)$ (see Fig.~\ref{Fitmoving}). The solid and dashed lines are obtained using the eigenenergies shown in Fig.~\ref{chiral_levels_movingframe} for the same level and $\vec{P}$.}\label{moving24L}
\end{figure}

To answer these question let us first illustrate the suppression at large $L$ of the second term in Eq.~(\ref{e}). To do this, we consider points in the energy region above the $\pi\Sigma$ threshold and close to the $\bar K N$ threshold for the different cases and determine the contribution of each of the two terms in Eq.~(\ref{e}).  In particular, for the symmetric box, we take the eigenenergies related to the level 1 shown in Fig.~\ref{ressym4}. In case of an asymmetric box, we use the points shown in Fig.~\ref{resasym}  for the level 1 with $z=0.5$ and for the system in a moving frame the two points in Fig.~\ref{Fitmoving} associated with the level 1 and the vector $\vec{P}=(2\pi/L)(2,0,0)$. We then use the $T_{11}$ matrix (i.e., $\pi\Sigma$ $T$-matrix) obtained for each case from the best fit to the energies shown in Figs.~\ref{ressym4}, \ref{resasym} and \ref{Fitmoving}, respectively, calculated at the corresponding eigenenergies. Next, for these eigenenergies, we determine $\delta G^{-1}_1$ and the difference $\delta G^{-1}_1-T_{11}$, which corresponds to the second term in Eq.~(\ref{e}). In Figs.~\ref{sym24L},~\ref{asym24L}, and ~\ref{moving24L} we show the results obtained for the symmetric box, asymmetric box and the system in a moving frame, respectively. In these figures, the filled triangles represent the modulus squared of the contribution arising from the L\"uscher approach, i.e., $\delta G^{-1}_1$,
while the empty triangles correspond to the contribution coming from the second term in Eq.~(\ref{e}). As we can see,  in case of a  symmetric box, the L\"uscher term in Eq.~(\ref{e}) dominates over the second one for $L\gtrsim 3.3m^{-1}_\pi$. For an asymmetric box, this happens for $L\gtrsim 3.8m^{-1}_\pi$.  In case of the system in a moving frame, since the biggest value of $L$ associated to the points considered in Fig.~\ref{Fitmoving}  is $L\sim2 m^{-1}_\pi$, to check if the second term of Eq.~(\ref{e}) is suppressed for bigger values of $L$ we have extended the calculation to higher values of $L$ by considering all the eigenenergies related to the level 1 and $\vec{P}=(2\pi/L)(2,0,0)$ of Fig.~\ref{chiral_levels_movingframe}.  The results are shown in Fig.~\ref{moving24L} as a solid line for the L\"uscher term and as a dashed line for the second term in Eq.~(\ref{e}). As can be seen, the value of $L$ after which the second term in Eq.~(\ref{e}) is negligible as compared to the first one is $L\gtrsim 2.7m^{-1}_\pi$. This value of $L$ is smaller than that found in a symmetric and an asymmetric box. Thus, it comes out that the use of a moving frame helps in reducing the volume dependence in the determination of phase shifts via the one channel Luescher formula, as shown in Ref.~\cite{arXiv:1108.5371}.

\begin{figure}
\includegraphics[width=0.45\textwidth]{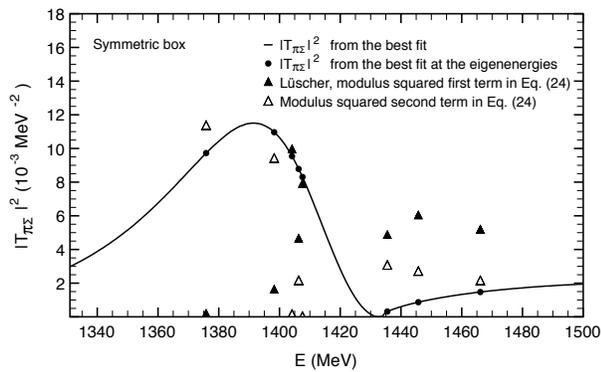}
\caption{Contributions arising from the first and second terms of Eq.~(\ref{e}) for the case of a symmetric box. The solid line corresponds to the modulus squared of the $\pi\Sigma$ $T$-matrix, $T_{\pi\Sigma}$, in infinite volume obtained from the best fit to the data points shown in Fig.~\ref{ressym4}. The circles correspond to the same quantity but evaluated at the data points (eigenenergies) of Fig.~\ref{ressym4} which fall above the $\pi\Sigma$ threshold and below 1500 MeV (a value higher than the $\bar K N$ threshold). The filled triangles represent the modulus squared of the Luescher term in Eq.~(\ref{e}) for the respective eigenenergies, while the empty triangles are for the contribution of the second term in Eq.~(\ref{e}).}\label{sym24}
\end{figure}

\begin{figure}
\includegraphics[width=0.45\textwidth]{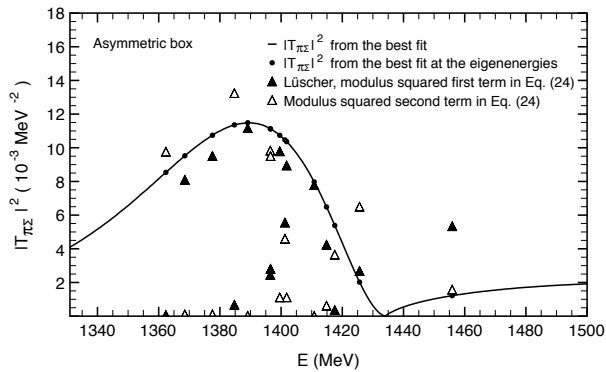}
\caption{Same as in Fig.~\ref{sym24} but for an asymmetric box (eigenenergies taken from Fig.~\ref{resasym}).}\label{asym24}
\end{figure}

\begin{figure}
\includegraphics[width=0.45\textwidth]{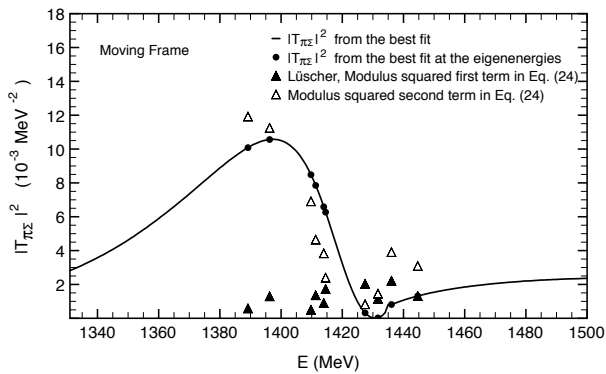}
\caption{Same as in Fig.~\ref{sym24} but for a moving frame (eigenenergies taken from Fig.~\ref{Fitmoving}).}\label{moving24}
\end{figure}

After showing the suppression with $L$ which occurs for the second term in Eq.~(\ref{e}), one may wonder why then one would need to consider coupled channels to determine
the poles of the $\Lambda(1405)$ and why L\"uscher approach would fail. One reason is that the channel $\pi \Sigma$ couples strongly to the lower pole and more weakly to the higher one, while the $\bar K N$ channel is the one that couples more strongly to the upper pole.  The $\bar K N \to \bar K N$  amplitude bears stronger information on the upper pole than the $\pi \Sigma \to \pi \Sigma$ one.  This second pole is also close to the $\bar K N$ threshold where the second term of Eq. (\ref{e}) will be important. We can see in practice the problems that we face in the realistic situation where the information is taken from the levels considered. Let us go back to the results shown in Figs.~\ref{sym24L},~\ref{asym24L},~\ref{moving24L}. There,
we have shown a suppression with $L$ for the second term in Eq.~(\ref{e}) with respect to the L\"uscher contribution to the scattering matrix for a certain energy level. However, it should be emphasized that the energy levels in the box start converging to some particular energy for large values of $L$ .  For example, the level 0 in Fig.~\ref{ressym4} starts from an energy of around 1280 MeV at $L=1.5\,m^{-1}_\pi$ and converges as $L$ increases to the threshold of the $\pi\Sigma$ channel. Similarly, the energy level 1 begins at an energy value close to 1370 MeV and as $L$ increases it stabilizes to an energy of around 1400 MeV, while the second level seems to converge to an energy around 1435 MeV, i.e., the $\bar K N$ threshold. Thus, by using a particular energy level in a range of $L$ one can not determine the scattering matrix for the system at infinite volume at any energy. To do this, we need to use different energy levels
covering different energy regions. For example, using the levels shown in Fig.~\ref{ressym4}, the scattering matrix in infinite volume can be determined at an energy of 1330 MeV using the level 0 at $L\sim4 m^{-1}_\pi$, but to obtain it at higher energies we need to shift to level 1, which starts at an energy of $~$1370 MeV at a much smaller value of $L$, $L=1.5 m^{-1}_\pi$ .
It would be interesting now to know what happens to the first and the second term in Eq.~(\ref{e}) when several energy levels, covering different energy ranges, are considered to determine the scattering matrix at infinite volume. We show the results obtained for the three cases studied, symmetric box, asymmetric box, and moving frame in Figs.~\ref{sym24}, ~\ref{asym24},~\ref{moving24}, using the eigenenergies  of Figs.~\ref{ressym4}, ~\ref{resasym},~\ref{Fitmoving}, respectively, which are above the $\pi\Sigma$ threshold (1331 MeV), up to a maximum energy of 1500 MeV (above the $\bar K N$ threshold). In these figures, the solid line corresponds to the modulus squared of the $\pi\Sigma$ amplitude in infinite volume obtained from the best fit to the energies shown in Figs.~\ref{ressym4},~\ref{resasym},~\ref{Fitmoving}. The circles correspond to the same quantity but for the respective eigenenergies which fall in the energy range 1331-1500 MeV. The filled and empty triangles represent the modulus squared of the first and second terms in Eq.~(\ref{e}), respectively, at the particular eigenenergies.  As one can see in these figures, there are large oscillations, and for a certain energy the L\"uscher term can be negligible as compared to the second one in Eq.~(\ref{e}), but for the next energy the situation can be completely reversed. Thus, the fact that one needs to use different energy levels to determine the scattering matrix in infinite volume below and above the different thresholds makes  it necessary to consider the contribution arising from the second term in Eq.~(\ref{e}). 

Once more, the exercise presented in this subsection reveals the relevance of the two channels in this particular problem and makes manifest the power of using effective theories at finite volumes as a prospective tool for further QCD lattice calculations.

\section{Conclusions}
   We have made a study of the $\bar K N$ interaction with its coupled channels in a finite box and found the levels obtained as a function of the box size. We have done this for standard periodic conditions and symmetric boxes, for asymmetric boxes and for symmetric boxes but with the particles in a moving frame. The aim of the work has been to solve the inverse problem in which, assuming that the levels in the box would correspond to ``QCD lattice results" we want to determine the pole positions in the complex plane for the two $\Lambda(1405)$ states provided by the chiral unitary approach and supported by several experiments. 
    
    We found that the problem is not trivial, and even the use of a large number of energies of the box corresponding to different levels and volumes with standard periodic conditions cannot provide the mass and width of the states with the accuracy of the chiral unitary approach and present experiments. For this reason we investigated other possible strategies and found that the use of asymmetric boxes and levels coming from the particles in moving frames helped considerably to narrow down the uncertainties in the determination of the mass and width of these resonances. The choices of levels and energies made for this analysis should be a guiding tool for future QCD lattice evaluations, showing the number of levels needed, the errors that should be demanded in the determination of the energies of the box and the type of asymmetric boxes or total momenta of the pair of particles in the moving frames. Having this information before hand is of tremendous value given the time consuming runs of actual QCD lattice runs.  
    
    Our analysis has also another important conclusion.  L\"uscher approach with one channel is universally accepted as an accurate tool to get phase shifts for energies where only one channel is open. In the present case we found that, since the  $\pi\Sigma$ and $\bar K N$ channels are not much separated in energy and are very entangled, the use of L\"uscher approach to get the $\pi\Sigma$ phase shifts from lattice energy levels leads to erroneous results, unless large volumes are used.  But one cannot anticipate how large should $L$ be to get a certain accuracy in the phase shifts. This should give us a warning for other cases and makes the finite volume studies within the chiral unitary approach very valuable as a prospective tool to be used for each individual case.

    We also found, in agreement with previous analytical studies, that the moving frames provide more accurate results for a given volume, via one channel analysis,  than the one at rest.

\section*{Acknowledgments}
 We would like to thank Michael D\"oring  for useful discussions. This work is partly supported by DGICYT contract FIS2011-28853-C02-01,
 the Generalitat Valenciana in the program Prometeo, 2009/090, and 
the EU Integrated Infrastructure Initiative Hadron Physics 3
Project under Grant Agreement no. 283286. This work is supported in part by
the Grant for Scientific Research (No.~24105706 and No.~22540275) from 
MEXT of Japan.
A part of this work was done in the Yukawa International Project for 
Quark-Hadron Sciences (YIPQS).
The work of A.~M.~T.~is supported by  
the Grant-in-Aid for the Global COE Program ``The Next Generation of Physics, 
Spun from Universality and Emergence" from the Ministry of Education, Culture, 
Sports, Science and Technology (MEXT) of Japan. M. Bayar acknowledges support through the Scientific and Technical
Research Council (TUBITAK) BIDEP-2219 grant.

\end{document}